\documentclass[aps,pre,twocolumn,showpacs,preprintnumbers,amsmath,amssym,superscriptaddress]{revtex4}
\usepackage{graphicx}
\usepackage{bm,color,textcomp}
\usepackage{amsmath,amsfonts,amssymb}

\usepackage{subfig}
\usepackage{graphicx}

\usepackage[normalem]{ulem}

\newcommand\mean[1]{\left<#1\right>}

\newcommand{\bx}{{\bf x}}
\newcommand{\ba}{{\bf a}}
\newcommand{\bb}{{\bf b}}
\newcommand{\bv}{{\bf v}}
\newcommand{\bu}{{\bf u}}
\newcommand{\bz}{{\bf z}}

\newcommand{\BE}{\begin{equation}}
\newcommand{\EE}{\end{equation}}
\newcommand{\BA}{\begin{eqnarray}}
\newcommand{\EA}{\end{eqnarray}}

\usepackage{soul} 

\begin{document}

\title{Cluster crystals with combined soft and hard-core repulsive interactions}
\author{Lorenzo Caprini}
\affiliation{Gran Sasso Science Institute (GSSI), Via. F. Crispi 7, 67100 L'Aquila, Italy.
}
\author{Emilio Hern\'andez-Garc\'\i a}
\affiliation{IFISC (CSIC-UIB), Instituto de F\'{\i}sica Interdisciplinar y Sistemas Complejos,
Campus Universitat de les Illes Balears, E-07122 Palma de Mallorca, Spain.
}

\author{Crist\'obal L\'opez}
\affiliation{IFISC (CSIC-UIB), Instituto de F\'{\i}sica Interdisciplinar y Sistemas Complejos,
Campus Universitat de les Illes Balears, E-07122 Palma de Mallorca, Spain.
}
\date{\today}

\begin{abstract}

Particle systems interacting with a soft repulsion, at thermal
equilibrium and under some circumstances, are known to form
cluster crystals, i.e. periodic arrangements of particle
aggregates. We study here how these states are modified by the
presence of an additional hard-core repulsion, accounting for
particle size. To this end we consider a two-dimensional system
of Brownian particles interacting through a potential which
includes a hard-core and a soft-core (of the GEM-$\alpha$ type)
repulsive terms. The system shows different phases, and we
focus in the regime where cluster crystals form. We consider
two situations: the low-temperature one in which particles
inside the clusters also show an ordered structure (crystal
cluster-crystal phase), and the one occurring at higher
temperature in which particles within the clusters are
spatially disordered (fluid cluster crystal). An explicit
expression for the energy in terms of the typical distance
between clusters and the typical distance of the particles
within the clusters is obtained for vanishing temperature, from
which mean mean inter- and intra-cluster distances are
obtained. Finite-temperature corrections are also discussed
considering explicitly the role of the entropy.

\end{abstract}

\maketitle

\section{Introduction}\label{sec:intro}

Particles interacting via soft-core repelling forces may
crystallize in equilibrium into ordered structures, with a unit
cell composed of a closely packed cluster of
particles~\cite{BookLikosSciortinoZaccarelli,Klein1994,
Likos2001, Mladek2006, Likos2007,
Mladek2008,CoslovichIkeda2013}. This is a particularly relevant
situation for some polymers, dendrimers or colloidal solutions,
where the effective potentials between their centers of mass
are soft and
repulsive~\cite{Likos2001a,Mladek2006,LikosPRL1998}. This
phase, called cluster crystal, occurs for low temperatures and
large enough density and packing fraction.

This type of systems has been studied by using a variety of
methods, from Monte Carlo to density functional
theory~\cite{Mladek2006,Likos2007}, or in the related
Dean-Kawasaki equation approach~\cite{Delfau2016}. A
mathematical criterion needed for the formation of cluster
crystals is that the Fourier transform of the interparticle
potential, $\tilde V (k)$, should take negative values for some
values of $k$~\cite{Mladek2006,Likos2007,Delfau2016}. For the
commonly studied  generalized exponential model (GEM-$\alpha$),
$V(r) = \epsilon \exp\left(-(r/R)^\alpha\right)$ -  which is an
example of soft potential~\cite{Likos2001a} - with $\epsilon$
an energy scale and $R$ the typical interaction range, this
happens when $\alpha>2$. The occurrence of cluster crystals in
GEM-$\alpha$ potentials with $\alpha>2$ is related to the fact
that as $\alpha$ increases the potential becomes more box-like
shaped. The physical mechanism leading to the cluster crystals
is the balance between the repulsive forces among particles
inside a cluster, and the forces from neighboring
ones~\cite{Delfau2016}.

The natural extension of considering a hard-core contribution
into the potential, taking into account the finite size and
impenetrable character of a core part of the interacting
particles, was considered in~\cite{Glaser2007}. A problem to
study this type of system within a continuous-density
description is that this approach leads to the consideration of
the Fourier transform of the potential, which is in general not
well-defined in the hard-core case. The authors
of~\cite{Glaser2007} use instead a lattice model where the
lattice constant equals the particle size and the hard-core
repulsion is implemented by imposing single occupancy per site.
Thus they generalize the soft-potential criterion for the
appearance of cluster crystals to include terms depending on
the occupied volume fraction. However, this lattice model
cannot distinguish ordered from disordered states within the
clusters. To avoid this  \cite{Glaser2007} also introduced an
off-lattice model with the particular type of potentials known
as hard-core/soft-shoulder potentials. Further studies with the
same potentials are found in~\cite{Shin2009, Ziherl2011} where,
in particular, finite temperature corrections and detailed
phase diagrams are presented.

In this work, we focus on the study of two-dimensional
off-lattice systems with repulsive interaction potentials
related to the well-studied hard-core/soft-shoulder case but
more general than it. Specifically, we consider a soft
repulsion of the GEM-$\alpha$ type at large distances,
complemented by a strongly repulsive core behaving as $r^{-b}$
at short distances. We keep the name of `hard-core' for this
part of the potential, although it is not strictly
impenetrable. Under suitable parameters, in particular at low
temperature, cluster-crystal phases are found. Our main
question is to understand how the hard-core power-law potential
changes the equilibrium configurations of particles interacting
through the GEM-$\alpha$ potential. To understand the
structures found we compute the energy of the system by
considering both the interactions between clusters and the
interactions among particles within clusters. For this last
contribution, at variance with previous
studies~\cite{Glaser2007,Shin2009, Ziherl2011}, we explicitly
take into account the intra-aggregate structure, i.e., whether
the particles may order periodically or not inside every
cluster. Mean inter- and intra-cluster distances are then
obtained from energy minimization. We validate our results with
numerical simulations of an ensemble of interacting Brownian
particles in a thermal bath. Temperature corrections are also
considered to estimate the typical cluster size, which is
different at very low temperature, when the within-cluster
distribution of the particles shows an ordered structure,  with
respect to higher temperatures, when the clusters are in a
fluid/gas state.



The paper is organized as follows. In Section~\ref{sec:model}
we introduce the model, discuss the role of the spatial scales
in the potential, and present the system's phases that will be
studied. In Section~\ref{sec:theory} we derive an analytical
expression for the energy of the system in the crystal
crystal-cluster phase, in the limit of small temperatures.   In
the subsequent section~\ref{sec:ft} we compute some temperature
effects on cluster characteristics. We conclude with a summary
and discussion in Sec.~\ref{sec:summary}.

\section{Model and cluster crystal phases}
\label{sec:model}

As one of the ways to obtain thermal equilibrium configurations
under the type of interactions described above, we consider a
$d$-dimensional system of $N$ interacting Brownian particles
of unit masses in contact with a thermal bath.
 We restrict to a regime of very large friction
which allows us to neglect the inertial terms and to assume
directly an overdamped dynamics:
\begin{equation}
\label{eq:model}
\dot{{\bf x}}_i = -\nabla_i V({\bf x}_1, ... , {\bf x}_N) +  \sqrt{2D}	\, \boldsymbol{\eta}_i, \ i=1,..,N,
\end{equation}
where the noise vector, $\boldsymbol{\eta}$, with component
$\eta_{l}$, verifies  $\left<\eta_{l}\right>=0$ and $\left<
\eta_{l}(t)\eta_{m}(t') \right>=\delta(t-t')\delta_{l,m}$, with
$l, m=1, ... , d$. The diffusion coefficient $D$ is
proportional to the temperature through the Einstein relation
$D= k_B T$ when time is measured in units of the inverse
friction coefficient.
The potential, $V$, is pairwise and
contains two parts, a hard-core repulsion referring to the
particle size, $V_h$, and a  soft-core repulsive part $V_s$
(they act at different scales as detailed later on):
\begin{equation}
V({\bf x}_1, ... , {\bf x}_N) =  \sum_{1\leq i<j \leq N} \big[ V_h(|{\bf x}_i -{\bf x}_j|)+V_s(|{\bf x}_i -{\bf x}_j|) \big].
\label{eq:potential}
\end{equation}
We restrict to $d=2$ and study all along this paper the
particular cases of
$V_h (r)=\epsilon_h (\frac{r_0}{r})^b$ and $V_s (r) =
\epsilon_s \exp {[-(r/R)^\alpha]}$, with $r=|\bf x|$. $V_s$ is
the GEM-$\alpha$ potential which accounts for an effective soft
repulsive interaction with a typical length scale $R$ and
energy scale $\epsilon_s$. $V_h$, modeling a hard-core, is a
standard power-law repulsion where $r_0$ stands for the typical
length scale for the size of the particles. $\epsilon_h$ is an
energy scale which can be absorbed in the definition of $r_0$,
since it always appears in the combination $\epsilon_h r_0^b$.
However, we keep this parameter explicit through the paper, as
it helps to identify the terms in our expressions coming form
the hard core. In the numerics it will be fixed as
$\epsilon_h=\epsilon_s$. Despite the particular choice of the
potentials we point out the generality of the arguments
presented below.

In the absence of the hard-core potential the behavior of the
system is well-understood: for $\alpha>2$, small enough
temperature, large enough numerical density, $\rho_0=N/L^d$,
and large packing fraction, $\phi=\rho_0 R^d$, cluster crystals
are formed in which (point) particles aggregate in clusters
separated by a distance proportional to $R$, with particles
randomly moving inside each cluster~\cite{Delfau2016}.
Unveiling the effects of the hard-core potential on this phase
is the main purpose in what follows. Note the difference with
other studies considering cluster phases with short-range
attraction and long-range
repulsion~\cite{SciortinoMossaPRL2004,MossaSciortinoTartagliaZaccarelli2004,ToledanoSciortino2009}.
In our setting, there is no attraction at any scale.

\begin{figure}[!h]
\includegraphics[width=0.8\linewidth,keepaspectratio]{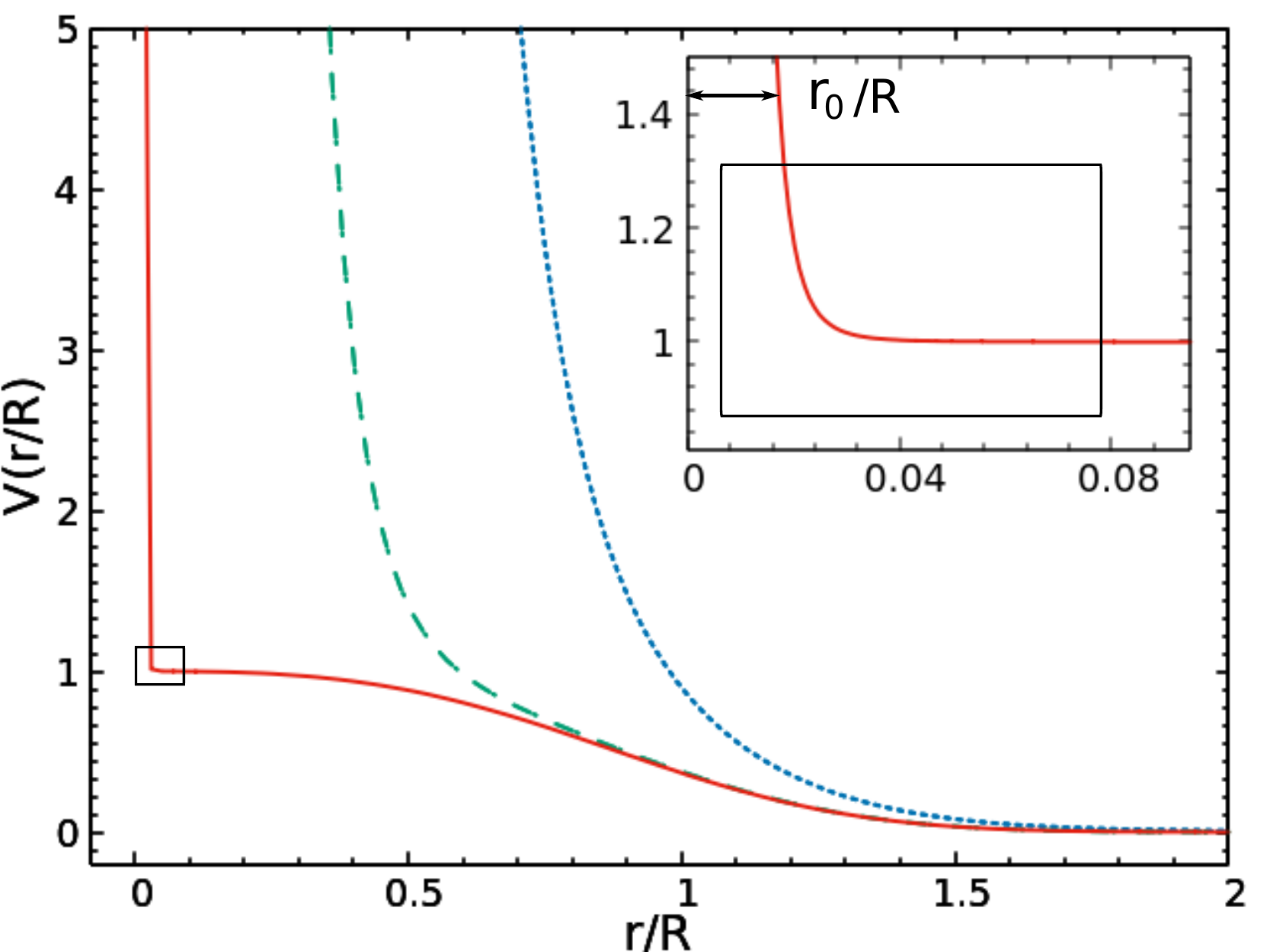}
\caption{(a) Potential given by $V_s(r)+V_h(x)$ with $V_s(r)=\epsilon_s\exp{\left(-(r/R)^{3}\right)}$
and $V_h(r)=\epsilon_h(r_0/r)^6$. Fixing $R=0.1$ and $\epsilon_h=\epsilon_s=1$, the different lines
correspond to: $r_0=0.001$ (red solid line), $0.04$ (green dashed line) and $0.1$ (blue dotted line).
In the inset, a zoom of the potential in the first case.}\label{fig:potential}
\end{figure}

The shape of the total potential $V = V_h + V_s$ is shown in
Fig.~\ref{fig:potential} for the special case $\alpha=3$ and
$b=6$. If there is no scale separation, $r_0 \gtrsim R$
(green-dashed and blue-dotted lines in
Fig.~\ref{fig:potential}), the potential behaves effectively as
just a hard-core repulsive one, since all of the features of
the soft-core part are masked. This is qualitatively similar to
setting $V_s=0$. In this case, particles freeze at low
temperatures forming a hexagonal lattice. We will not discuss
here the subtleties associated with the nature of this type of
two-dimensional crystallization, see for
example~\cite{Dudalov2014}. The interesting regime, assumed in
the rest of the paper, occurs when $r_0 \ll R$ (red solid line
in Fig.~\ref{fig:potential}): the potential has an abrupt
decrease up to the scale $r_0$, followed by a
smoother one, as shown in the figure inset.

\subsection{Cluster crystal phases}
\label{sec:pd}

As shown in Fig.~\ref{fig:behavior}, numerical simulations in a
two-dimensional square box of size $L$ with periodic boundary
conditions show a phenomenology consistent with~\cite{Glaser2007}.
In particular, for $\alpha =3$, $b=6$ and restricting to
densities and packing fractions large enough, cluster crystal
phases are found: at low-enough temperature, particles
aggregate in clusters which arrange hexagonally. Moreover, the
average population of each cluster, $N_c$, increases with
$\phi$. The inter-cluster distance, $\bar{x}$, is relatively
insensitive to $\phi$, increasing just with $R$. These are the
usual properties of cluster
crystals~\cite{Mladek2006,Likos2007,Delfau2016}.

We discuss the effect of the temperature $T$, or equivalently
of the diffusion coefficient $D=k_B T$, by evaluating
the microscopic structure of the system via the radial
distribution function $g(r) = \rho_0^{-1} \langle\sum_{i \neq
0} \delta (\bx -  \bx_i)\rangle$, where a target particle is at
the origin, the sum is over the other particles, and the
brackets indicate an equilibrium temporal (long-time) or
thermal average, and also a circular average over positions
$\bf x$ with the same modulus $| {\bf x}|=r$.

The comparison of $g(r)$ at small distances with typical radial
distribution functions of a simple solid/liquid/gas allows us
to identify solid-, liquid- and gas-like aggregation phases
inside the clusters, which coexist with the cluster crystal
structure at much larger distances. Since the number of
particles in the clusters is small, these states cannot be
interpreted as real distinct macroscopic phases. For this
reason, a rigorous discussion about phase transitions is
meaningless and we restrict our considerations to the
qualitative configuration of the system, for which the
following observations are made:
\begin{figure}[!h]
\includegraphics[width=0.95\linewidth,keepaspectratio]{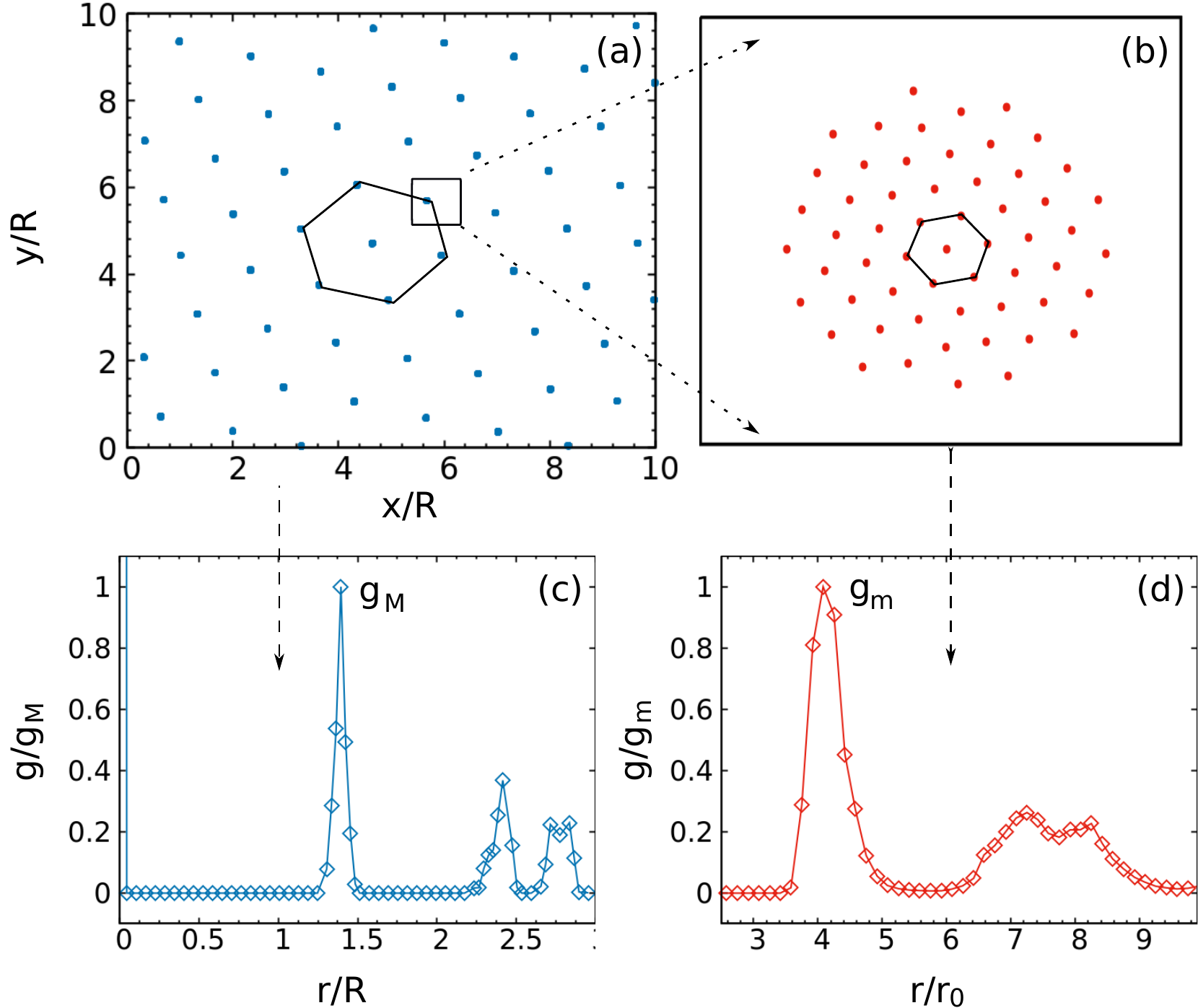}
\caption{Crystal cluster-crystal phase.
Top panels: A snapshot of the system configuration in the $x$-$y$ plane (Panel (a)) and zoom into
a single cluster (Panel (b)).
Bottom panels: Radial correlation function $g(r)$ as a function of $r/R$ (Panel (c)). This representation highlights
the presence of a peak at $r\approx 1.4 R$, indicating periodic ordering with this periodicity.
Also visible is a huge peak at $r\approx 0$, indicative of particle clustering at small distances.
A zoom of $g(r)$ on this peak is plotted in Panel (d), as a function of $r/r_0$. This part of
$g(r)$ characterizes the type of particle ordering inside the clusters.
For presentation purposes, $g$ is normalized by the height of the main peak in each graph.
Parameters: $\alpha =3$, $b=6$, $L=1$, $N=1000$, $r_0=10^{-3}$,
$R=10^{-1}$, $\epsilon_s=\epsilon_h=1$, $D=10^{-6}$.
}
\label{fig:behavior}
\end{figure}

\begin{figure}[!t]
\includegraphics[width=0.95\linewidth,keepaspectratio]{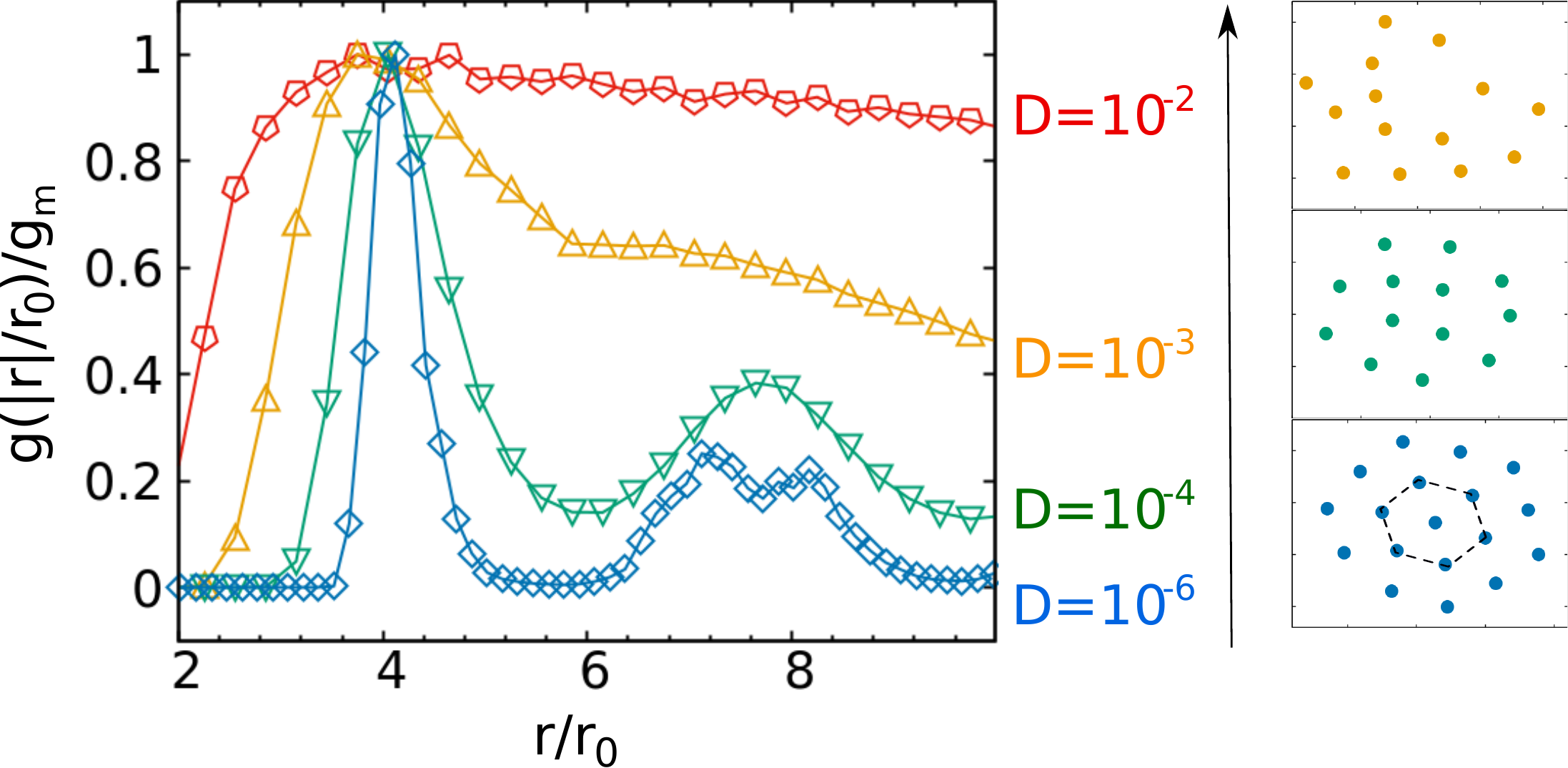}
\caption{$g(r)/g_m$ ($g_m$ is the maximum value of $g$) at small scales $r\approx r_0<< R$
for different diffusion coefficients $D=k_B T=10^{-6}, 10^{4}, 10^{-3}, 10^{-2}$, respectively,
blue diamonds, green upper triangles, orange lower triangles and red pentagons.
On  the right we show a snapshot of within-cluster particle distributions corresponding to each of the three lower temperatures.
Remaining parameters:  $\alpha=3$, $b=6$, $L=1$, $N=10^3$, $r_0=10^{-3}$, $R=10^{-1}$, $\epsilon_s=\epsilon_h=1$}
\label{fig:temp}
\end{figure}

\begin{enumerate}
\item For small temperatures, particles inside each cluster
    are almost frozen,  showing a hexagonal arrangement.
    This is the ordered crystal phase identified in
    \cite{Glaser2007}, which is shown in the top row of
    Fig. \ref{fig:behavior}. The left panel shows the
    crystal of clusters, and the right one a zoom of one
    cluster, with the particles perfectly ordered. The
    lower row shows $g(r)$ at the scale of the clusters
    (left panel) and a zoom into the peak in the region
    $r\approx r_0$, displaying the crystalline ordering of
    the particles within clusters (right panel). From the
    position of the peaks in both figures, we can obtain
    the equilibrium distance among clusters $\bar{x}_e
    \approx 1.4 R$ (left panel) or among particles
    $\hat{x}_e\approx 4.05 r_0$ (right). As in a solid,
    particles fluctuate around their equilibrium positions,
    but these fluctuations are very small compared to
    $\hat{x}_e$.  We observe that in the steady state,
    after a long time transient, the hexagonal clusters
    become aligned so that the lines joining their centers
   coincide with their apothems: the crystal is completely
   ordered.

\item Increasing the temperature, particle positions
    disorganize and
a liquid-like phase appears inside each cluster. As in a
simple liquid, particles move around the whole cluster
volume. In Fig.~\ref{fig:temp}, $g(r)$ at the small scales
is shown (green $D=10^{-4}$ and yellow $D=10^{-3}$ lines),
revealing a crossover from solid to fluid characteristics.

\item A further increase of $T$ produces the growth of
    the clusters volume. Particles within each cluster form
    a gas-like phase, as revealed by the absence of
well-defined secondary peaks in Fig.~\ref{fig:temp} (red
$D=10^{-2}$ line).

\item Further increase of the diffusion coefficient or
    temperature makes the clusters disappear, leaving a
    global fluid phase. On the other hand, by increasing
    $\phi$, we observe the lamellar and micelle phases
    described in~\cite{Glaser2007}. However, as mentioned
    in the introduction, we restrict our study to the range
    of temperatures or volume fractions where only the
    different types of cluster crystals appear.
\end{enumerate}

\section{Cluster energy}\label{sec:theory}

We now set-up an energy calculation which is useful to
understand the observed structures and phenomenology, in
particular, the steady state properties of the system: the
distance between clusters, the lattice type and the average
size of the clusters. In this Section we restrict to very small
$T$, i.e., to the crystal cluster-crystal phase, numerically
discussed in the previous Section. In this regime any entropic
contribution does not affect significatively the free energy,
as is expected  in the solid state. In the next Section we will
consider finite temperature corrections, and the role of the
entropy will be explicitely taken into account.
Note that, as mentioned in the introduction, we avoid any
description based on the coarse-grained density of particles.
This is similar to the off-lattice approach followed by~\cite{Glaser2007} for the hard-core/soft-shoulder case, but
extended here to our more general class of potentials.


We rewrite the total energy into two contributions: a
self-energy of the clusters themselves, $E_C$, and an
interaction energy among them, $E_I$.
Two main quantities are fundamental: the mean
inter-cluster distance, $\bar x$, and the mean inter-particle
distance inside a cluster, $\hat x$. Indeed, the number of
clusters in the system, $n_c$, can be estimated as $n_c=\chi
L^d/\bar{x}^d $, (for arbitrary dimension $d$) where $\chi$ is
 a parameter which depends on the lattice, for
instance $\chi=1$ for a square lattice and $\chi= 2/\sqrt{3}$
for a triangular one. If we denote as $N_c$ the average number
of particles of each cluster, then $N  = n_c N_c$, so that we
can write $N_c = \bar{x}^d\rho_0 /\chi$.
The average size of the cluster, $\sigma_c$, can be  estimated from $\hat{x}$ and $N_c$, considering the
hexagonal arrangement of the intra-cluster particles.

In the following, we work in $2$ dimensions  but our approach can be  extended to three dimensional systems.
As mentioned, for small enough
temperature we can neglect entropic contributions and the
equilibrium properties are all determined by the total energy
$E = E_C + E_I$, where:
\begin{equation}
E_C= n_c  \sum_{1\leq i<k }^{N_c} \left[ V_h(|{\bf x}_i-{\bf x}_k|)+V_s(|{\bf x}_i-{\bf x}_k|)\right].
\end{equation}
Since $r_0$ is very small, and $V_s$ decays fast with distance,
it is a reasonable approximation for the intercluster energy to
consider only interactions of particles in neighboring
clusters, so that:
\begin{flalign}
\label{eq:Eh}
E_I & \approx  n_c \frac{\gamma_I}{2} E^{a,b}_I   \\
&=n_c \frac{\gamma_I}{2} \sum_{k_b, i_a=1}^{N_c}  \left[V_s(|{\bf x}_{i_a}- {\bf x}_{k_b}|) + V_h(|{\bf x}_{i_a}- {\bf x}_{k_b}|)    \right],\nonumber
\end{flalign}
where $E_I^{a,b}$ is the interaction energy between two
neighboring clusters, that we denote by $a$ and $b$, and ${\bf
x}_{i_a}$ ($i_a=1,...,N_c$) is the position of the $i$-th
particle in the  cluster $a$. The constant $\gamma_I$ is the
number of first neighbors in the lattice in which the clusters
arrange (in two dimensions $\gamma_I=6$ for a hexagonal
lattice, or $\gamma_I=4$ for a square lattice). Since we assume
a separation of scales of both potential terms, we consider
$\bar{x}$ and $\hat{x}$ independent variables and look for an
expression for the energy such that $E = E(\bar{x}, \hat{x}) =
E_I(\bar{x}, \hat{x}) + E_C(\bar{x}, \hat{x}) $. As already
mentioned, for $T$ small enough, the equilibrium configuration
is the one that minimizes the total energy with respect to
$\bar{x}$ and $\hat{x}$. Let us point out that, because of the
relation between $\bar{x}$ and $N_c$, at constant density the
minimization with respect to $\bar{x}$ is formally equivalent
to the one with respect to $N_c$, exploited within the density
functional approach in previous works \cite{Mladek2008multiple,
Likos2008} for purely soft-core potentials.

To proceed we introduce the particular forms of $V_h$ and $V_s$ and make use of their
properties. In the following we are labeling different clusters
with letters $a,b,...=1,2,...,n_c$ and the particles with
indices $i,j,...=1,...,N$ or if we want to specify the $i$-th
particle  of cluster $a$ we use $i_a=1,...,N_c$. Let us
first work out the self-interaction cluster energy:
\begin{eqnarray}
\label{eq:Es0}
E_C &=& n_c \sum_{i<k}^{N_c}\left[\epsilon_s e^{-(|{\bf x}_i-{\bf x}_k|/R)^{\alpha}} +
\epsilon_h \left(\frac{r_0}{|{\bf x}_i-{\bf x}_k|}  \right)^b  \right] \nonumber \\
&= & n_c\epsilon_s
 \left(\frac{N_c}{2}(N_c-1)+\sum_{i<k}^{N_c} {\cal O}(x_{ik}/R)  \right) \nonumber \\
& + & n_c \epsilon_h \sum_{i<k}^{N_c} \left(\frac{r_0}{x_{ik}}\right)^b \nonumber \\
&\approx & n_c\epsilon_s \frac{N_c}{2}(N_c-1) +  n_c \epsilon_h N_c \gamma_C \left(\frac{r_0}{\hat{x}}  \right)^b,
\end{eqnarray}
where $x_{ik}$ is the distance between the $i$-th and $k$-th
particle in one cluster, so that $|x_{ik}|<d_c$, where $d_c$ is
the diameter of the cluster, which is also smaller than the
inter-cluster distance, $d_c< \bar{x}$. The last approximation
in Eq.~\eqref{eq:Es0} is obtained by considering  $|x_{ik}|/R
\ll 1$, and also restricting interactions with the hard-core
potential to the first neighbors.
Now, the constant $\gamma_C$ is the number of first neighbors
in the intra-cluster lattice, ($\gamma_C=6$ in the hexagonal
case). $E_C$ in Eq.~\eqref{eq:Es0} contains a term coming from
the soft-core part of the potential, and another involving the
hard-core one, which takes into account the internal structure
of the clusters. Using $N=n_c N_c$ and $N_c =
\bar{x}^2\rho_0 /\chi$, we find (neglecting a constant term):
\begin{equation}
E_C = N \left(\epsilon_s \frac{\rho_0}{\chi} \frac{\bar{x}^2}{2} +
\epsilon_h \gamma_C \left(\frac{r_0}{\hat{x}}  \right)^b  \right).
\label{eq:Es}
\end{equation}
We can see from this expression that the cluster self-energy
favors the equilibrium configuration where clusters are near
(because of the term $\propto \bar{x}^2$), but with particles
inside them far from each other (because of the term proportional to
$1/\hat{x}^b$).

Let us next obtain an expression for the interaction energy
among first neighbors clusters, $E_I$, starting from
Eq.~\eqref{eq:Eh}. The distance between two particles in
different clusters can be written as $|{\bf x}_{i_a} - {\bf
x}_{k_b}| = |(\bar{x}+z_{i_a k_b}){\bf \hat{z}} + y_{i_a
k_b}{\bf \hat{y}}|$, where $\bar{x}+z_{i_a k_b}$ is the
distance between the particles of two different clusters
projected onto the direction of the unit vector ${\bf \hat{z}}$
parallel to the line connecting cluster centers; $y_{i_a k_b}$
is the projection of the same distance on the orthogonal axis
of unit vector ${\bf \hat{y}}$. In this way we have separated
the distance between two particles in different clusters as a
sum of the distance between the centers of the two clusters,
$\bar x$, and the remaining part $z_{i_a k_b}$, which is
smaller than the diameter of the cluster $d_c$. Taking into
account the specific shape of the potentials:
\begin{eqnarray}
\label{eq:ssp}
&E_I  =  n_c \frac{\gamma_I}{2} \sum_{k_b, i_a=1}^{N_c} \Biggl[\epsilon_s
e^{-(|(\bar{x}+z_{i_a k_b}){\bf \hat{z}} + y_{i_a k_b}{\bf \hat{y}}|/R)^{\alpha}}& \nonumber \\
& + \epsilon_h \left(\frac{r_0}{|(\bar{x}+z_{i_a k_b}) {\bf \hat{z}} + y_{i_a k_b}{\bf \hat{y}}| } \right)^b  \Biggr]& \nonumber \\
 &\approx   n_c \frac{\gamma_I}{2} \sum_{k_b,i_a=1}^{N_c}
\epsilon_s e^{-\left(|(\bar{x}+z_{i_a k_b}){\bf \hat{z}} + y_{i_a k_b}{\bf \hat{y}}|/R\right)^{\alpha}},&
\end{eqnarray}
where we have neglected the term ${\cal O}(r_0/|(\bar{x}+z_{i_a
k_b}){\bf \hat{z}} + y_{i_a k_b}{\bf \hat{y}}|)\ll 1$, since
$\bar{x}\gg r_0$. The role of the hard-core is present in this
expression through $z_{i_a k_b}$ and $y_{i_a k_b}$ in the
exponential, which is not negligible and gives a non-trivial
contribution.  Note that taking into account the microscopic
structure inside a cluster is crucial for the micro-scale: the
appearance of $z_{i_a k_b}$ and $y_{i_a k_b}$ in Eq.
(\ref{eq:ssp}) gives the only dependence on $\hat{x}$ that can
balance the $1/\hat{x}^b$ term appearing in the cluster
self-energy Eq. (\ref{eq:Es}). If this dependence on $\hat{x}$
is neglected, it would be impossible to find any equilibrium
value, $\hat{x}_e$.

\begin{figure}[!t]
\includegraphics[width=1\linewidth,keepaspectratio]{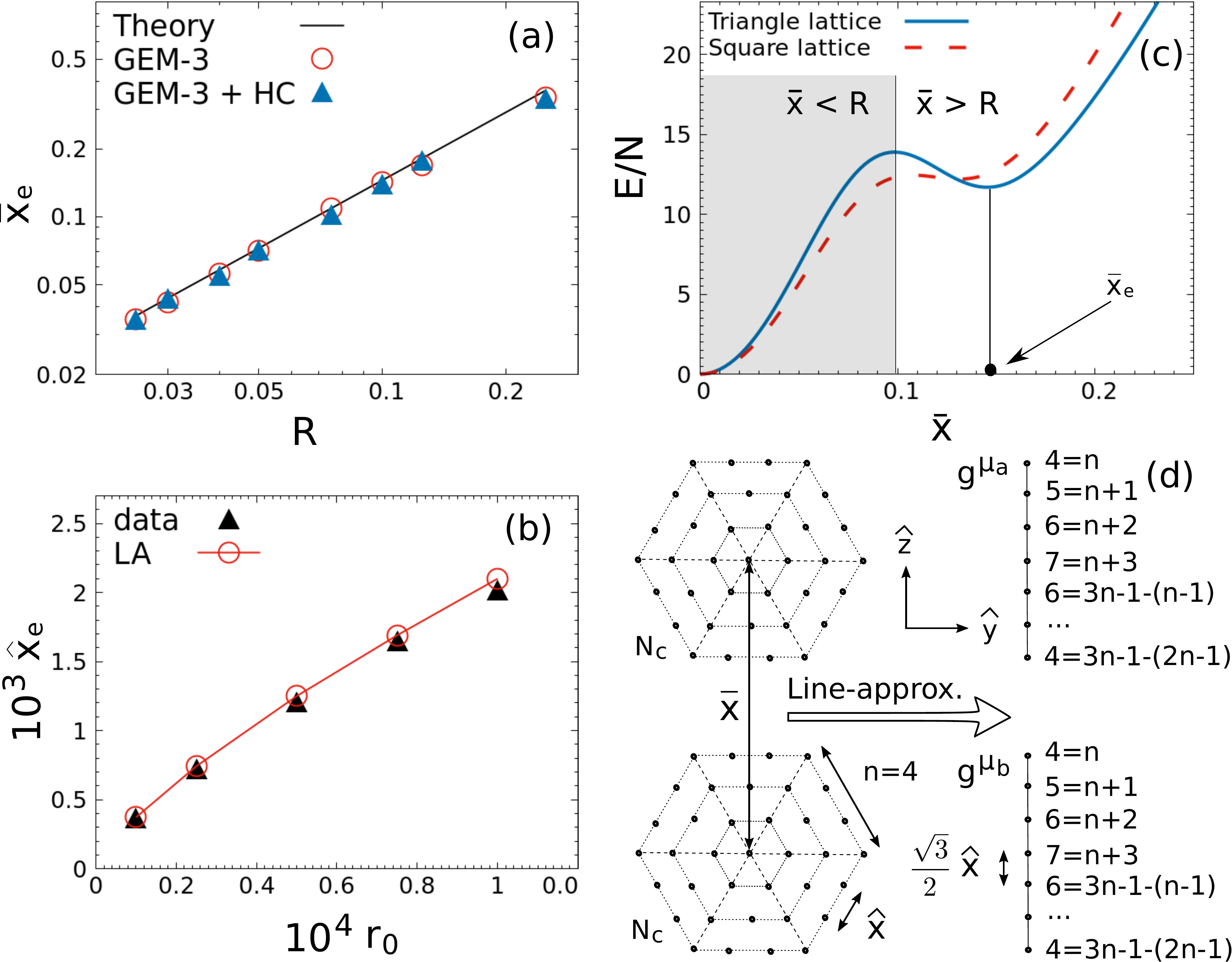}
\caption{
Left Panels: $\hat{x}_e$ and $\bar{x}_e$ versus $R$ (panel (a)) and $r_0$ (Panel (b)),
obtained from the simulation of the particle dynamics (symbols) and from minimization of
expression~\eqref{eq:final}
for $E(\bar{x}, \hat{x})$ (lines), assuming hexagonal intracluster and intercluster lattices.
The theoretical line in Panel (a) is indistinguishable from the line 
$\bar x_e \approx 1.45 R$ resulting from minimizing Eq. (9) neglecting 
the hard core.
Panel (c): $E(\bar{x},\hat{x}_e)$
vs $\bar{x}$ at the minimum of the intracluster particle distance $\hat{x}_e$ (with $\gamma_C=6$),
for a square intercluster lattice (red dashed line) and a hexagonal one (blue solid line).
The black point stems from $\bar{x}_e$ from the Brownian simulations, which coincides with the
lowest energy minimum, which occurs for the hexagonal configuration. As discussed
in the text, the dashed region $\bar{x}<R$, and then the additional minimum occurring at $\bar{x}=0$,
should be disregarded since our expressions are not valid there.
Panel (d): graphical illustration of the line approximation (LA), discussed in the text, for
a pair of contiguous clusters with $n=4$. Numbers in the right give the values of the degeneracy factors
$g^{\mu_a}$ and $g^{\mu_b}$ counting the number of particles projected by the LA onto the same site in
the vertical line.
Parameters: $D=10^{-6}$, $r_0/R=10^{-2}$, $\epsilon_s=\epsilon_h=1$, $L=1$, $N=10^3$, $\alpha=3$, $b=6$. }
\label{fig:check}
\end{figure}

To close our problem we need to  write the distance
$|(\bar{x}+z_{i_a k_b}){\bf \hat{z}} + y_{i_a k_b}{\bf
\hat{y}}|$ in terms of $\hat{x}$ and $\bar{x}$,  in order to express $E_I$ in terms of these two distances.
One of the possible approximations,  which we call line
approximation (LA), consists in neglecting the $y$ coordinate
of the particles, so that each  hexagonal cluster becomes
compressed onto a line. Since, as noted above, the direction
$z$ joining cluster centers coincides with the hexagonal
cluster apothems, the LA considers all particles concentrated
on these apothems. See Fig. \ref{fig:check}, panel d), for a
drawing of this procedure. By calling $n$ the number of
particles located on one external side of a hexagonal cluster,
the approximation projects particle positions into one of the
$2n-1$ sites, separated by a distance $\hat{x}\sqrt{3}/2$,
located in a line along the $z$ direction (see Fig.
\ref{fig:check}d). Each site, now, contains more than one
particle, with a degeneracy which is maximum in the middle of
the cluster and minimum at its extremal points.
$n$ is related to $N_c$ by: $N_c= 3n (n-1) + 1$. Since
$N_c=\bar{x}^2 \rho_0/\chi$, we can express $n$ as a function
of $\bar{x}$. For large hexagons $n\gg1$, we have $n\approx
\bar{x}\left(\rho_0/3\chi\right)^{1/2}$. Using that $n_c=N/N_c = N/\chi \rho_0 \bar{x}^2$, we can easily find an approximation for
$E_I$:
\begin{equation}
\label{eq:Eh_final}
E_I\approx   \epsilon_s N \frac{\chi}{\rho_0}\frac{1}{\bar{x}^2} \gamma_I
\sum_{\mu_a,\mu_b=1}^{2n-1} g^{\mu_a} g^{\mu_b} e^{ -\left(\left|\bar{x} +
(\mu_a+\mu_b-2n) \frac{\sqrt{3}}{2} \hat{x}\right|/R\right)^\alpha},
\end{equation}
where the sum is over all sites, $\mu_a$ and $\mu_b$, in the
lines that approximate the particle positions in two contiguous
clusters. The product $g^{\mu_a} g^{\mu_b}$ gives the
degeneracy arising from the number of particles associated to
each site in each of the two lines: $g^\nu=n-1+\nu$ for $\nu
\in[1,n]$ and $g^\nu=3n-1-\nu$ for $\nu \in[n+1, 2n-1]$ (Fig.
\ref{fig:check}d).

Unlike the expression for the cluster self-energy,
Eq.~\eqref{eq:Es}, $E_I$ favors the configurations for which
the mean cluster distance $\bar{x}$ is large but $\hat{x}$ (the
intracluster particle distance) is small, as will be seen
later.

To better display the dependencies, we normalize
both the sum in Eq.\eqref{eq:Eh_final} (which scales as $\sim
n^2$) and the degeneracy, $g^{\mu_a}$, introducing
$\tilde{g}^{\mu_a} =g^{\mu_a}/n$. Considering the previous
relation between $n$ and $\bar{x}$, and combining
Eq.\eqref{eq:Es} and Eq.\eqref{eq:Eh_final}:
\begin{eqnarray}
&&\frac{E(\bar{x}, \hat{x})}{N} =
\epsilon_s \frac{\rho_0}{\chi} \frac{\bar{x}^2}{2} +
\epsilon_h \gamma_C \left(\frac{r_0}{\hat{x}}  \right)^b   \nonumber \\
&+&\epsilon_s  \frac{\rho_0\, \bar{x}^2}{9} \frac{\gamma_I}{\chi} \frac{1}{n^2}
\sum_{\mu_a,\mu_b=1}^{2n-1} \tilde{g}^{\mu_a} \tilde{g}^{\mu_b} e^{ -\left(\left|\bar{x} + (\mu_a+\mu_b-2n)\frac{\sqrt{3}}{2} \hat{x}\right|/R\right)^\alpha}.
\label{eq:final}
\end{eqnarray}
Eq.~\eqref{eq:final} is our main analytical result. It contains
three terms, which help us to understand the different effects
giving rise to the crystal cluster-crystal phase arising at
very low temperatures. The first term describes the tendency of
the soft potential to favor clusters as close as possible
(small $\bar{x}$), simply since in this way, at constant mean
density $\rho_0$, each cluster would be less populated and the
internal repulsion will be smaller. This tendency is opposed by
the third term, which comes also from the soft potential, and
contains coupled geometric contributions from both $\bar{x}$
and $\hat{x}$. With respect to the $\bar{x}$ dependence the third term in Eq.(\ref{eq:final})
is approximatively
$\propto \bar{x}^2
\exp{\left(-\left(\bar{x}/R\right)^{\alpha}\right)}$, if we
neglect the weak influence of $\hat{x}$. The interplay between
this dependence of the third term and the $\bar{x}^2$
dependence of the first one singles out two values of $\bar x$
by energy minimization: $\bar{x}_m=0$ and $\bar{x}_e>R$. The
first one, and in general the values coming from Eq.
Eq.~\eqref{eq:final} for $\bar{x} < R$, are not reliable within
the approximations used: when $\bar{x} < R$ interactions with
clusters beyond nearest neighbors would need to be taken into
account, which would raise the energy. Also, the condition of
having a cluster lattice without empty clusters implies
$\bar{x}\geq l \approx L/\sqrt{N}$. Indeed if $\bar{x}= l$ the
clusters have population equal to one ($N_c=1$) and the cluster
self-energy is equal to zero and thus exploring $\bar{x}$ below
this limit is meaningless. Thus, neglecting the behavior for
$\bar{x}<R$, the interplay between the first and third terms
selects at zero temperature an equilibrium value $\bar{x}>R$
which we call $\bar{x}_e$ (see Fig.~\ref{fig:check}~(a) and
(c)). $\bar{x}_e$ is nearly insensitive to the value of
$\hat{x}$.

The second term in Eq.~\eqref{eq:final} arises from the
hard-core potential and simply expresses that this short-range
repulsion favors large $\hat{x}$. Again, the third term
balances this tendency since
larger $\hat{x}$ imply that some particles of different
clusters come closer (the particles such that $\mu_a + \mu_b - 2n < 0$), which is unfavorable for the soft-core
intercluster repulsion. The balance between the two tendencies
determines the zero-temperature equilibrium value of $\hat{x}$,
which we call $\hat{x}_e$. Taking into account the
$\hat{x}$-dependence in Eq. \eqref{eq:Eh_final} is crucial to
predict a finite $\hat{x}_e$. Indeed, this is the only
dependence which can balance the repulsive $1/\hat{x}^b$ term
appearing in the cluster self-energy. We also note that if
$\epsilon_h=0$ all particles in a cluster would collapse to its
center ($\hat{x}=0$), which is indeed what happens at zero
temperature in the absence of the hard-core potential.

In order to check our approximate expression, in
Fig.~\ref{fig:check} we plot as functions of $R$ and $r_0$ the
equilibrium intercluster distance, $\bar{x}_e$, and the
equilibrium inter-particle distance inside a cluster,
$\hat{x}_e$, as obtained from numerical simulations of the
Brownian set of particles, showing a good agreement with the
theoretical prediction from the minimization of our energy
expression. In particular, in the case of the intercluster
distance (Panel (a) of Fig.~\ref{fig:check}) we compare
the prediction with two different simulation settings, with and
without the hard-core potential (purely soft-core potential).
As expected the smaller scale does not influence much the
intercluster distance $\bar{x}_e$, which is quite constant with
respect to $r_0$, for $r_0\ll R$. This confirms, in agreement
with~\cite{Glaser2007}, that the microscopic details of the
clusters are macroscopically less relevant.
The relevance of the small-scale appears in determining the
intracluster distance $\hat{x}_e$ (Fig.~\ref{fig:check} panel (b)), which depends strongly on $r_0$, and so the
average size of the clusters (which is roughly
$\hat{x}_e\sqrt{N_c}$).


The panel (c) of Fig.~\ref{fig:check} plots
$E(\bar{x},\hat{x}=\hat{x}_e)$ as a function of the
intercluster distance $\bar{x}$ for parameters $\chi$ and
$\gamma_I$ corresponding to two different cluster lattices
(hexagonal and square). It is seen that the hexagonal minimum
is lower (remember than the minimum at $\bar{x}=0$ and the
whole region $\bar{x}<R$ should be disregarded as
Eq.~\eqref{eq:final} is not valid there), implying that the
hexagonal cluster crystal is the most stable. Similar results
appear in~\cite{Shin2009} for the hard-core/soft-shoulder
potential in the context of the lattice description.



\section{Finite temperature corrections }\label{sec:ft}

The theoretical results of the previous section are
approximately valid only in the limit of vanishing temperature.
At finite temperature, energy minimization should be replaced
by free-energy minimization, which requires to take into
account the role of the entropy, $ S$. Approaches in this line
for the lattice shoulder potential can be found
in~\cite{Glaser2007,Shin2009}. In this Section, we estimate the
free energy and use it to characterize temperature effects on
the cluster structure.

In principle, we can distinguish between different entropic
sources: the inter-cluster entropy, $S_I$, which takes into
account the defects of the cluster-crystal lattice, and the
internal entropy of the clusters, $S_C$. In general, $S_I \ll
S_C$, if $N_C \gg 1$ since, roughly, $S_I \propto n_c = N/N_c$
and $S_C \propto N$. For this reason, we focus on the
calculation of $S_C\approx S$, which strongly depends on the
intra-cluster regime we are considering. Below, we compute it
in the crystal cluster-crystal and in the fluid (gas-like)
cluster-crystal phases, where we developed reasonable
approximations for the probability distribution of the particle
positions. In an intermediate liquid cluster-crystal regime,
the situation is different since all the intra-cluster
particles have positions correlated in a non-trivial way, as
usual in the framework of liquid theory. We point out that in
the fluid cluster-crystal regime $S_C$ plays a fundamental role
in determining the average size of the cluster, in agreement
with \cite{Delfau2016} for purely soft-core potentials.

\subsection{Very low temperature: Crystal cluster-crystal phase}
\label{subsec:cccp}

We consider first the situation of very low temperature, in
which the cluster crystal with crystalline interior remains,
and neglect any temperature influence in the positions of the
cluster centers (which then remain in a hexagonal lattice of
intercluster distance $\bar x_e$).
In the same way, we assume that particles inside each cluster
remain close to the positions in the hexagonal lattice of
distance $\hat x_e$ which characterizes the zero-temperature
equilibrium state. This implies neglecting any thermal dilation
effect which could affect $\bar x_e$ or $\hat x_e$. The only
temperature effect we estimate now is the possible vibration of
each particle around its equilibrium position, characterized by
a vibration width $\sigma_h$.
To this end, we concentrate on the vibrational entropy $S_v$,
neglecting any \emph{translational} entropy associated with the
displacement of the equilibrium particle positions, which will
eventually favor the formation of defects.


We make a further mean-field-like approximation in order to
evaluate $S_v$: interactions among particles are already taken
into account in determining the lattice constants $\bar x_e$
and $\hat x_e$ and we neglect any further effect in producing
particle-position correlations. More specifically, we assume
the $N$-particle probability of positions to factorize:
$P_N({\bf x}_1,...,{\bf x}_N)=\prod_{a=1}^{n_c}
\prod_{i_a=1}^{N_c} p({\bf x}_{i_a}-{\bf u}_{i_a})$, where the
first product runs over the $n_c$ clusters and the second over
$\{{\bf u}_{i_a}\}$, the (zero-temperature) equilibrium
positions of the particles. $p({\bf x})$ gives the probability
distribution of the position fluctuations of each particle
around its equilibrium. We consider the same $p({\bf x})$ for
each particle, independently of its position in the cluster.
Then, the entropy associated to $P_N$ becomes a sum of terms,
one for each particle:
\begin{equation}
S_v= n_c N_c S_1 \ , \ S_1=- k_B \int d{\bf x} p({\bf x}) \log \left(p({\bf x})/\rho_0\right) \ .
\end{equation}
Our strong-localization assumption implies that no
combinatorial factor coming from indistinguishability needs to
be used. Our final assumption is that at very low temperatures
$p({\bf x})$ is a narrow two-dimensional Gaussian of width
$\sigma_h$ \cite{tarazona1985free}, so that:
\begin{eqnarray}
S_v &=& - k_B n_c N_c \int d{\bf x} \frac{e^{-{\bf x}^2/2\sigma_h^2}}{2\pi\sigma_h^2}
\log\left( \frac{e^{-{\bf x}^2/2\sigma_h^2}}{2\pi\sigma_h^2 \rho_0}\right) \nonumber \\
&=& k_B n_c N_c \left( 1+\log(2\pi)+\log(\sigma_h^2 \rho_0) \right) \ .
\label{eq:Sv_lowT}
\end{eqnarray}
We have used that for a two-dimensional isotropic Gaussian
variable, $\mean{\bx^2}=\mean{x^2}+\mean{y^2}=2\sigma_h^2$. The
width $\sigma_h$ is the only remaining parameter and will be
determined from the minimization of the free-energy. To be
consistent with the approximations used, the resulting width
should satisfy $\sigma_h \ll \hat{x}$.

Next, we evaluate the average energy of the system, separated
in intracluster and intercluster interactions:
\begin{equation}
\begin{aligned}
\langle E \rangle&= \langle E_C \rangle + \langle E_I \rangle \\
&= \sum_{a=1}^{n_c} \sum^{N_c}_{i_a<j_a}
\int d\{\bx\}_N P_N(\bx_1,...,\bx_N) V(\bx_{i_a} - \bx_{j_a}) \\
&+ \sum_{a=1}^{n_c} \sum_{a<b}^{n_c}\sum_{i_a,j_b}^{N_c}
\int d\{\bx\}_N P_N(\bx_1,...,\bx_N) V(\bx_{i_a} - \bx_{j_b}) .
\label{eq:fullenergy}
\end{aligned}
\end{equation}
Under the same approximate mean-field framework as for the
entropy, we split $P_N$ into a product of single-particle
(Gaussian) probabilities. For the intracluster part we arrive
to:
\begin{flalign}
&\mean{E_C} \approx \nonumber\\
&n_c \frac{N_c}{2} \int d\bx_1 d\bx_2
\frac{e^{-\bx_1^2/2\sigma_h^2}}{2\pi\sigma_h^2} \frac{e^{-\bx_2^2/2\sigma_h^2}}{2\pi\sigma_h^2}
\sum_{j\ne 0} V(\bx_1- \ba_j - \bx_2) \nonumber\\
&= n_c \frac{N_c}{2} \int d\bz
\frac{e^{-\bz^2/4\sigma_h^2}}{4\pi\sigma_h^2} \sum_{j\ne 0}V(\bz- \ba_j) \ .
\label{eq:Es_lowT}
\end{flalign}
The sum is over all position vectors $\{\ba_j\}$ of the
equilibrium positions of the particles inside a cluster (except
the one at $\ba_0={\bf 0}$, where we have arbitrarily located
the equilibrium position of the first particle), which form a
hexagonal lattice. The last equality is obtained after changing
variables to the average, $\bu=(\bx_1+\bx_2)/2$, and relative,
$\bz=\bx_1-\bx_2$, coordinates, and integrating over $\bu$.

The interaction potential is made of the hard-core and the
soft-core parts, $V(\bx)=V_h(|\bx|)+V_s(|\bx|)$. Since the
Gaussian restricts the integration to a region of size
$\sigma_h$ around the origin, and we are assuming $R \gg \hat
x_e \gg \sigma_h$, the soft-core potential is effectively
constant inside the integral, $V_s(|\bx|)=\epsilon_s (1+{\cal
O}(\sigma_h/R)^2)$ and then $\sum_{j\ne 0}V_s(|\bv-
\ba_j|)\approx (N_c-1)\epsilon_s$. For the hard-core potential
we approximate the interaction sum by the contribution from the
nearest neighbors of the particle at the origin, which are at
the corners of a hexagon ($\ba_1,...,\ba_6$, with $|\ba_i|=\hat
x_e$) (see upper right panel of Fig. \ref{fig:behavior}).
Expanding the interaction sum in the vicinity of $\bz={\bf 0}$:
\BE
\sum_{i=1}^6 V_h(|\bz- \ba_i|) \approx 6 V_h(\hat x_e)
+\frac{3}{2} \left( V_h(\hat x_e)'' + \frac{V_h(\hat
x_e)'}{\hat x_e}\right) \bz^2 + ...
 \label{eq:hexagonal_expansion_h}
\EE
The terms neglected are of order $z_x^4$, $z_y^4$ and $z_x^2
z_y^2$, and thus will give corrections smaller than
$(\sigma_h/\hat x_e)^4$. Introducing in Eq.~\eqref{eq:Es_lowT}
and performing the integration:
\begin{flalign}
&\mean{E_C} \approx \\
& n_c \frac{N_c}{2} \left( (N_c-1)\epsilon_s + 6 V_h(\hat x_e)
+ 6 \left( V_h(\hat x_e)'' + \frac{V_h(\hat x_e)'}{\hat
x_e}\right) \sigma_h^2 \right) \nonumber\ .
\end{flalign}

Considering now the intracluster part of the mean energy, we
again factorize the $N$-particle probability. In the previous
section, the contribution of the particle positions inside the
cluster to $E_I$ was needed to properly determine the
interparticle distance $\hat{x}$. But here, once we take this as
fixed, we estimate the temperature corrections to $\mean{E_I}$
by assuming that all particle equilibrium positions are at the
center of the cluster they belong. Under this approximation:
\begin{flalign}
\label{eq:Eh_lowT}
&\mean{E_I} \approx  \\
&n_c \frac{N_c^2}{2} \int d\bx_1 d\bx_2
\frac{e^{-\bx_1^2/2\sigma_h^2}}{2\pi\sigma_h^2}
\frac{e^{-\bx_2^2/2\sigma_h^2}}{2\pi\sigma_h^2} \sum_{b\ne 0}
V(\bx_1- \bb_b - \bx_2) \nonumber\ .
\end{flalign}
The sum is now over all position vectors $\{\bb_b\}$ of the
cluster centers (except the one we arbitrarily locate at
$\bb_0={\bf 0}$), which form again a hexagonal lattice.

Since $r_0\ll \bar x_e=\min_{b\ne 0}\{|\bb_b|\}$, the
contribution of the hard-core potential to $\mean{E_I}$ is
negligible. As before, except for corrections which are ${\cal
O}(\sigma_h/R)^2$ the soft-core potential in each term of the
sum can be considered constant inside the integral. Restricting
the sum to the six clusters surrounding (at distance $\hat
x_e$) the one at the origin, we find:
\BE
\mean{E_I} \approx n_c \frac{N_c^2}{2} 6 V_s(\bar x_e) \ .
\EE

We can now write the full expression for the free energy:
\begin{equation}
\label{eq:F_hardcore_temperaturecorrection}
\begin{aligned}
\frac{F}{n_c N_c}&= \frac{\mean{E} - T S}{n_c N_c} \approx
\frac{E_0}{n_c N_c} - D (1+\log(2\pi)) \\
&+ 3 \left( V_h(\hat x_e)'' + \frac{V_h(\hat x_e)'}{\hat x_e}\right)
\sigma_h^2 -D \log(\sigma_h^2\rho_0) \ .
\end{aligned}
\end{equation}
Where  we have used $D=k_B T$,
and we have collected all terms of the mean energy which are
independent of $\sigma_h$ into the constant $E_0$.
By minimizing with respect to $\sigma_h$ and taking the
explicit forms for $V_s$ and $V_h$ we find:
\BE
\sigma_h^2 = \frac{D}{3 \left( V_h(\hat x_e)'' + V_h(\hat
x_e)'/\hat x_e\right)} = \frac{D \hat x_e^2 (\hat x_e/r_0)^b}{3
b^2 \epsilon_h} \ , \label{eq:sigma_h}
\EE
which is the expression characterizing the influence of
temperature on the fluctuations of each particle position
around its equilibrium location. Note that this is precisely
the expression for the standard deviation of the Gaussian
probability which describes the motion of a Brownian particle
in the harmonic approximation, close to the origin, due to the
combined potential of six particles at hexagonal positions
surrounding the origin at distance $\hat x_e$. The consistency
condition $\sigma_h\ll \hat x_e$ implies that
Eq.~\eqref{eq:sigma_h} is only valid at small temperatures such
that $D \ll D_h=3 b^2 \epsilon_h (r_0/\hat x_e)^b$. One may
think that $D_h$ gives a rough estimation of the transition
temperature above which the crystalline structure inside the
clusters is lost (probably into a liquid-like phase), but we
think it is at best a rough upper bound, because of the many
approximations involved. In a liquid-like phase particles do
not fluctuate around any equilibrium points. Moreover, the
dynamics is correlated meaning that the main hypothesis of our
calculation is not satisfied. Minimizing Eq.
\eqref{eq:F_hardcore_temperaturecorrection} also with respect
to the intra- and inter-cluster distances we can obtain
entropic corrections to them. These corrections are ${\cal
O}(D)$ and therefore very small in this phase.



\subsection{Fluid cluster-crystal phase}
\label{subsec:fccp}

Numerical simulations indicate that there is a range of
temperatures in which clusters remain but particles inside them
do not display a crystal structure, but a fluid-like behavior.
This implies that thermal motion inside the cluster has
exceeded the capacity of the hard-core potential to keep the
particles in place, as it would occur if $D\gtrsim D_h$.
Because of this, and since we know that clusters appear because
of the nature of the soft-core repulsion (forming a large
hexagonal lattice with lattice vectors $\{\bb_b\}$), we
describe this fluid-cluster crystal state by completely
neglecting the hard-core potential, i.e. $V(\bx)\approx
V_s(|\bx|)$. In this gas-like situation, we now estimate how
the cluster width $\sigma_s$ depends on temperature.

In the same mean-field approach as before, we consider that the
many-body probability, $P_N$, factorizes into single-particle
Gaussians - but this time of width $\sigma_s$ - characterizing
cluster size, since each particle can explore the whole cluster
in the fluid state. A consistency condition is that $\sigma_s
\ll \bar x_e$ for the crystal-cluster structure to remain
despite the finite size of the clusters. In fact, we should
also have $\hat x_e \sqrt{N_c}\ll\sigma_s$, since the first
term is an estimation of the size of a cluster of $N_c$
particles in the low-temperature regime within which it retains
a crystal structure. Similarly to the previous low-temperature
case (see Eq.~\eqref{eq:Sv_lowT}), the vibrational entropy will
be:
\BE
S_v = k_B n_c N_c \left( 1+\log(2\pi)+\log(\sigma_s^2 \rho_0)
\right) \ .
\label{eq:SV_highT}
\EE
For the mean energy, we distinguish again the cluster
self-energy $\mean{E_C}$ and the intercluster contribution
$\mean{E_I}$. For this last quantity, we introduce the
probability factorization into Eq.~\eqref{eq:fullenergy} to
obtain (c.f. Eq.~\eqref{eq:Eh_lowT}):
\BA
&&\mean{E_I} \approx  \nonumber \\
&& \frac{n_cN_c^2}{2} \int d\bx_1 d\bx_2
\frac{e^{-\bx_1^2/2\sigma_s^2}}{2\pi\sigma_s^2}
\frac{e^{-\bx_2^2/2\sigma_s^2}}{2\pi\sigma_s^2} \sum_{b=1}^6
V_s(|\bx_1- \bb_b - \bx_2|) \nonumber \\
&&= \frac{n_cN_c^2}{2} \int d\bz
\frac{e^{-\bz^2/4\sigma_s^2}}{4\pi\sigma_s^2} \sum_{b=1}^6
V_s(|\bz- \bb_b|) \ .
\EA
We have approximated all equilibrium particle positions as
located at the center of the cluster they belong, and
interactions have been restricted to the six clusters (with
$|\bb_b|=\bar x_e$) neighboring the first one, which we have
arbitrarily located at the origin. The last equality is
obtained after transforming to relative and center-of-mass
coordinates and integrating over the last one. Using the
expansion (c.f. Eq.~\eqref{eq:hexagonal_expansion_h})
\BE
\sum_{b=1}^6 V_s(|\bz- \bb_b|) \approx 6 V_s(\bar x_e)
+\frac{3}{2} \left( V_s(\bar x_e)'' + \frac{V_s(\bar
x_e)'}{\bar x_e}\right) \bz^2 + ...\ ,
 \label{eq:hexagonal_expansion_s}
\EE
we obtain:
\BE
\mean{E_I} \approx 3 n_c N_c^2 \left( V_s(\bar x_e) + \left(
V_s(\bar x_e)'' + \frac{V_s(\bar x_e)'}{\bar x_e}\right)
\sigma_s^2 \right)  \ . \label{eq:Eh_highT}
\EE
The intracluster self-energy reads:
\BA
&&\mean{E_C} \approx \nonumber \\
&&n_c\frac{N_c}{2} \int d\bx_1 d\bx_j
\frac{e^{-\bx_1^2/2\sigma_s^2}}{2\pi\sigma_s^2}
\frac{e^{-\bx_2^2/2\sigma_s^2}}{2\pi\sigma_s^2}
\sum_{j \ne 1}^{N_c} V_s(|\bx_1 - \bx_j|) \nonumber \\
&&=n_c \frac{N_c (N_c-1)}{2} \int d\bz
\frac{e^{-\bz^2/4\sigma_s^2}}{4\pi\sigma_s^2} V_s(|\bz|) \ .
\EA
For the GEM-$\alpha$ potential, $ V_s(|\bz|) =
\epsilon_s(1+{\cal O} (|\bz/R|^\alpha))$, being the last term
negligible, if $\alpha>2$, compared to the terms already
considered in Eq. (\ref{eq:hexagonal_expansion_s}). This is
precisely the reason why cluster crystals form in a
GEM-$\alpha$ potential with $\alpha>2$: the particle repulsion
inside the cluster is negligible compared with the repulsion
from the neighboring clusters \cite{Delfau2016}. Thus,
neglecting terms smaller than ${\cal O} (|\sigma_s/R|^2))$:
\BE
\mean{E_C} \approx n_c \frac{N_c (N_c-1)}{2} \epsilon_s \ .
\EE
The complete expression for the free energy in this fluid or
gas cluster regime is:
\begin{equation}
\label{eq:F_TemperatureCorrections}
\begin{aligned}
&\frac{F}{n_c N_c}= \frac{\mean{E} - T S}{n_c N_c} \approx
\frac{\tilde E_0}{n_c N_c} - D (1+\log(2\pi))\\
&\quad+ 3 N_c \left( V_s(\bar x_e)'' + \frac{V_s(\bar x_e)'}{\bar x_e}\right) \sigma_s^2 -D \log(\sigma_s^2\rho_0)  \ .
\end{aligned}
\end{equation}
All terms of the mean energy which are independent of
$\sigma_s$ have been included in $\tilde E_0$. Minimizing with
respect to $\sigma_s$ and taking our explicit expression for
$V_s$ gives:
\BE
\begin{aligned}
\sigma_s^2 &= \frac{D}{3 N_c \left( V_s(\bar x_e)'' + V_s(\bar
x_e)'/\bar x_e\right)} \\
&= \frac{D \bar x_e^2 e^{(\bar
x_e/R)^\alpha}(R/\bar x_e)^\alpha}{3\alpha^2\epsilon_s N_c
\left((\bar x_e/R)^\alpha -1 \right)}  \ . \label{eq:sigma_s}
\end{aligned}
\EE
As in Eq.~\eqref{eq:sigma_h}, this is the width of the Gaussian
probability distribution for a Brownian particle moving in the
combined potential of six clusters located on the corners of a
hexagon at distance $\bar x_e$ from the origin, each providing
a repulsion given by the potential $N_c V_s$, and under the
harmonic approximation close to the origin. It also coincides,
after noting that the number of particles in each cluster of a
hexagonal cluster crystal is $N_c=\rho_0\sqrt{3}\bar x_e^2/2$,
with the cluster width derived from approximations to the
Dean-Kawasaki equation in a GEM-$\alpha$
potential~\cite{Delfau2016}. We recall that
expression~\eqref{eq:sigma_s} is expected to be valid only in
the intermediate temperature range such that $\hat
x_e\sqrt{N_c} \ll \sigma_s \ll \bar x_e$. Violation of the last
inequality, i.e. $\sigma_s\approx
\bar x_e$, gives a rough upper limit to the melting temperature
of the cluster crystal. The results in~\cite{Delfau2016}
indicate that indeed this estimation overestimates the cluster
crystal melting temperature (for the system in which only the
interaction $V_s$ is present) although the qualitative
parameter dependence is correct. As a final remark, minimizing
Eq.\eqref{eq:F_TemperatureCorrections} with respect to the
inter-cluster distance would allow us to estimate the
temperature corrections in determining $\bar{x}_e$. These
corrections are all ${\cal O}(D)$.


\section{Summary and discussion}
\label{sec:summary}

We have studied the influence of a hard-core potential on the
cluster crystal phase of a system of particles interacting
through a GEM-$\alpha$ repulsive potential. Performing
off-lattice numerical simulations of the interacting Brownian
particles we have identified the different ordering types that
particles inside the clusters exhibit. Temperature drives a
transition from a crystal cluster-crystal scenario (where
particles within clusters are periodically ordered) to a
fluid/gas cluster-crystal one. In the small temperature limit,
 an energy expression has been obtained
which helps to understand the balances between the different
forces leading to the existence of the cluster-crystal phase.
In addition, the finite temperature value of the cluster width
has been obtained for the fluid cluster-crystal state, and also
for the fluctuation amplitude of the particles around their
equilibrium positions in the crystal cluster-crystal state.
They provide rough upper limits to the temperatures at which
transitions would take place.

The methodology employed is rather general, and other types of
potentials, leading to cluster crystals, could be considered.
See for instance the soft potential
in~\cite{Cinti2014,Diaz-Mendez2015} in the context of bosonic
interactions. Furthermore, we expect our approach to be of use
to describe biological aggregations in which the individuals
have a finite size and interact through forces acting
attractively and/or repulsively at different scales or through
competing/mutualistic dynamics~\cite{Khalil2017}. The
generalization of our study to non-equilibrium systems of
particles with finite-size and interacting through repulsive
forces is of much interest in the context of active matter and
will be considered in the future, extending approaches such as
those in Ref. \cite{DelfauLopezGarcia2017}.

\section{Acknowledgements}

We acknowledge financial support from the Spanish grants LAOP
CTM2015-66407-P (AEI/FEDER, EU) and ESOTECOS
FIS2015-63628-C2-1-R (AEI/FEDER, EU). We acknowledge fruitful
discussions and continuous support from Profs. Umberto Marini
Bettolo Marconi and Angelo Vulpiani.

\bibliography{bibliodef}



\end{document}